\documentclass[usenatbib,useAMS]{mnras}
\pdfoutput=1
\usepackage{natbib}
\usepackage{amsmath}
\usepackage{graphicx}
\usepackage{color}
\usepackage{mathrsfs}
\usepackage{epsfig}
\usepackage{comment}
\usepackage{ulem}

\newcommand{\msun}{{\rm M}_\odot}

\newcommand{\zsun}{Z_\odot}

\newcommand{\cc}{{\rm cm}^{-3}}

\newcommand{\msunyr}{{\rm M}_\odot~{\rm yr}^{-1}}

\newcommand{\ckpc}{{\rm ckpc}}
\newcommand{\mpc}{{\rm Mpc}}
\newcommand{\cmpc}{{\rm cMpc}}
\newcommand{\gpc}{{\rm Gpc}}
\newcommand{\pc}{{\rm pc}}
\newcommand{\kms}{{\rm km~s}^{-1}}

\newcommand{\K}{{\rm K}}
\newcommand{\beq}{\begin{equation}}
\newcommand{\eeq}{\end{equation}}

\defcitealias{K14}{K14}
\defcitealias{K16}{K16}
\defcitealias{Visbal_2015}{VHB15}

\title[Massive seeds \& PopIII galaxies]
{Massive black hole and Population III galaxy formation in 
over-massive dark matter halos with violent merger histories}

\author[]{Kohei Inayoshi$^1$
\thanks{E-mail: inayoshi@astro.columbia.edu}
\thanks{Simons Society of Fellows, Junior Fellow},
Miao Li$^2$ and Zolt\'an Haiman$^1$
\\
$^{1}$Department of Astronomy, Columbia University, 550 W. 120th Street, New York, NY 10027, USA,\\
$^{2}$Center for Computational Astrophysics, Flatiron Institute, 162 Fifth Avenue, New York, NY 10010, USA
}

\begin{document}
\maketitle
\label{firstpage}

\begin{abstract}
We propose the formation of massive pristine dark-matter (DM) halos with masses of 
$\sim 10^8~\msun$, due to the dynamical effects of frequent mergers in rare 
regions of the Universe with high baryonic streaming velocity relative to DM.
Since the streaming motion prevents gas collapse into DM halos and delays prior star formation episodes,
the gas remains metal-free until the halo virial temperatures $\ga 2\times 10^4~\K$.
The minimum cooling mass of DM halos is boosted by a factor of $\sim 10-30$
because frequent major mergers of halos further inhibit gas collapse.
We use Monte Carlo merger trees to simulate the DM assembly history 
under a streaming velocity of twice the root-mean-square value,
and estimate the number density of massive DM halos containing pristine gas as 
$\simeq 10^{-4}~\cmpc^{-3}$.
When the gas infall begins, efficient Ly$\alpha$ cooling drives cold streams penetrating 
inside the halo and feeding a central galactic disc.
When one stream collides with the disc, strong shock forms a dense and hot gas cloud, 
where the gas never forms H$_2$ molecules due to effective collisional dissociation.
As a result, a massive gas cloud forms by gravitational instability and collapses
directly into a massive black hole (BH) with $M_\bullet \sim 10^5~\msun$.
Almost simultaneously, a galaxy with $M_{\star, \rm tot}\sim 10^6~\msun$ composed of 
Population III stars forms in the nuclear region.
If the typical stellar mass is as high as $\sim 100~\msun$, the galaxy could be detected 
with the {\it James Webb Space Telescope} even at $z\ga 15$.
These massive seed BHs would be fed by continuous gas accretion from the host galaxy, and grow to be 
bright quasars observed at $z\ga 6$.
\end{abstract}


%%%%%%%%%
%	Section 1	%
%%%%%%%%%
\section{Introduction}

The origin of supermassive black holes (SMBHs),
which are ubiquitous at the centers of galaxies, is one of the most challenging puzzles in astrophysics.
Observations in recent decades have revealed that SMBHs with masses of $M_\bullet \ga 10^{9}-10^{10}~\msun$ 
have already formed only within one billion years after the Big Bang 
\citep{Fan_2006,Willott_2010,Mortlock_2011,Wu_2015,Jiang_2016,Matsuoka_2016,Matsuoka_2018,Banados_2018}.
The existence of such monster SMBHs in the early Universe requires rapid formation and growth processes of BHs
\citep[][and reference therein]{Volonteri_2012_Review,Haiman_2013_Review,Johnson_2016_Review}.

In this paper, we consider the formation of massive seed BHs
\citep{Loeb_&_Rasio_1994, Eisenstein_Loeb_1995,Oh_Haiman_2002,Koushiappas_2004,
BVR_2006,Lodato_&_Natarajan_2006,Mayer_2010}, where 
a supermassive star (SMS) with $\ga 10^5~\msun$ formed in a protogalaxy 
collapses directly into a massive BH seed via general relativistic instability 
\citep{Shibata_&_Shapiro_2002,Montero_2012,Uchida_2017}.
SMS formation can be achieved in an H$_2$-free pristine gas cloud with an
almost constant and high temperature of $\sim 10^4~\K$.
Such a self-gravitating cloud collapses into a single star or several clumps due to fragmentation
at the center of the halo \citep{Regan_2014, IOT14, Becerra_2015, Latif_2016, Chon_2018}.
Most of massive clumps would migrate inward and merge rapidly due to gravitational interaction 
with the surrounding gas disc \citep{IH_2014,Sakurai_Vorobyov_2016}.
Since the hot gas envelope accretes on the central protostar at a high rate of 
\begin{equation}
\dot{M}\sim \frac{c_{\rm s}^3}{G}\simeq 0.18~\msunyr
\left(\frac{T}{10^4~\K}\right)^{3/2},
\end{equation}
\label{eq:acc}
\citep{Larson_1969,Penston_1969,Shu_1977},
the star can grow to an SMS with $M_\star \ga 10^5~\msun $ within the lifetime of $\sim 1$ Myr.
Because of rapid entropy input by accretion, the protostar has a bloated stellar envelope with
an effective temperature of $\sim 5000~\K$, which is much lower than that of normal massive stars
\citep{Hosokawa_2012}.
Thus, the accreting protostar is unlikely to suffer either from radiation feedback or from pulsation-driven mass loss,
and eventually collapses into a massive BH with $M_\bullet \ga 10^5~\msun$
\citep{IHO_2013,Hosokawa_2013,Umeda_2016,Haemmerle_2017}.
Such a massive seed could grow up to $\ga 10^9~\msun$ 
through subsequent gas accretion by $z\ga 6-7$ \citep[e.g.,][]{DiMatteo_2012,Smidt_2017}.

So far, H$_2$ dissociation by Lyman-Werner (LW) photons emitted from nearby star-forming galaxies 
has been mainly discussed in order to form massive seed BHs.
In this scenario, intense LW radiation is required to prevent H$_2$ formation 
\citep[e.g.,][]{O01,BL_2003, SBH_2010, IO11,WHB_2011,Johnson_2013,Latif_2014,
Sugimura_2014,Agarwal_2015,Chon_2016,WHB_2017}.
The critical LW intensity (in units of $10^{-21}$ erg s$^{-1}$ cm$^{-2}$ Hz$^{-1}$) is estimated as
$J_{\rm LW,21}\sim 10^3$, assuming that radiation sources have the same spectra 
as in metal-poor galaxies \citep{Sugimura_2014}.
\citealt{WHB_2017} have improved the estimate considering optically-thick H$_2$ photo-dissociation rates
over all the LW lines self-consistently with the spectra of source galaxies, though most other work just 
adopted a shielding factor instead.
As a natural outcome, such star-forming galaxies are also likely to produce intense X-rays due to the existence of 
high-mass X-ray binaries and supernova explosions.
However, X-ray ionization promotes H$_2$ formation through the electron-catalyzed reactions 
\citep*{Haiman_Rees_Loeb_1996},
and thus the critical LW intensity for H$_2$ dissociation is boosted 
\citep{IO11,Latif_2015,Inayoshi_Tanaka_2015,Glover_2016}.
As a result of this, the X-ray emission could invalidate the formation of massive seed BHs 
in the majority of halos exposed to LW radiation with $J_{\rm LW,21}>10^3$.

Alternatively, H$_2$ collisional dissociation (${\rm H} + {\rm H}_2 \rightarrow  3{\rm H}$)
potentially plays an important role in the formation of massive seed BHs 
in dense and hot shock regions satisfying 
\begin{equation}
\left( \frac{n}{10^4~\cc}\right) \left(\frac{T}{10^4~\K}\right )\ga 1
\label{eq:ZONR_1}
\end{equation}
\citep{IO12},where $n$ and $T$ are the density and temperature of the shock gas.
The shocked gas is expected to form by colliding accretion flows in the assembly of 
atomic-cooling halos with virial temperature of $T_{\rm vir}\ga 8000~\K$.
However, \cite{Fernandez_2014} found that for halos with masses and virial temperatures
near the atomic-cooling threshold ($\sim 10^4~\K$), 
cold gas flows accrete into the halo but experience shocks before reaching the central region.
In this case, the shocked gas is not dense enough to dissociate H$_2$.
This happens because the dynamical timescale of the flow is comparable to the radiative cooling 
timescale ($t_{\rm dyn} \sim t_{\rm cool}$).
\cite{visbal_14} have confirmed that dense shocked gas never forms without radiative cooling 
($t_{\rm dyn} \ll t_{\rm cool}$)\footnote{One possible idea proposed by \cite*{Inayoshi_2015} 
is a high-velocity collision of two DM haloes with a relative velocity $\ga 200~\kms$, 
inside which star formation has not occurred yet because of lower densities ($n\sim 10^2~\cc$). 
The shocked gas due to such a violent galaxy collision heats up to $\sim 10^6~\K$ and cools isobarically 
via free-free emission and He/H lines, resulting in dense gas with $n\ga 10^4~\cc$ at $T\sim 10^4~\K$.}.
Therefore, we require {\it more massive} haloes with $T_{\rm vir}\ga 10^4~\K$,
where the shock-dissipated energy is quickly carried away by radiative cooling.
Since the atomic hydrogen cooling rate $\Lambda_{\rm H}$ depends steeply on the gas temperature 
($\Lambda_{\rm H} \propto T^\beta$ where $\beta \sim 8$ at $8000 \la T/\K \la 2\times 10^4$, 
and $\Lambda_{\rm H}$ has a peak value at $T\simeq 2\times 10^4~\K$) ,
one can expect that the gas properties in massive DM halos with $T_{\rm vir}\ga 10^4~\K$ 
would change significantly due to radiative cooling. 
Analytical models and numerical simulations for lower-redshift and more massive galaxies 
with $T_{\rm vir}\gg 10^4~\K$ suggest that cold streams can penetrate inside the halo and 
form shocks near the center
(\citealt{Birnboim_Dekel_2003,Dekel_Birnboim_2006,Dekel_2009}, see also e.g.,
\citealt{Rees_Ostriker_1977,White_Rees_1978,White_Frenk_1991}).

One important concern is that such massive halos with $T_{\rm vir}> 10^4~\K$
would be likely polluted by metals due to stellar activity in their progenitors.
Metal pollution allows additional cooling by C/O lines and dust emission,
leading to efficient gas fragmentation and suppressing SMS formation 
\citep{OSH_2008,IO12,Latif_Omukai_2016}\footnote{Alternatively, \cite{Mayer_2015} have proposed that
in a very massive halo with $M_{\rm vir}\simeq 10^{12}~\msun$ at $z\sim 10$ ($T_{\rm vir}\sim 10^7~\K$),
shock heating driven by intense inflows could prevent gas fragmentation even in the face of the metal-line cooling.}.
To avoid metal pollution in massive halos, we here consider the effect of baryonic 
streaming motion (BSM) induced at the cosmological recombination epoch \citep{TH_2010}.
The BSM is a natural mechanism suppressing gas collapse into halos and 
delaying formation of the first generation stars (Population III, hereafter PopIII)
in mini-halos at $z>15$ \citep{Greif_2011, Stacy_2011,Fialkov_2012}. 
Applying the effect of high streaming velocities in semi-analytical merger-tree calculations,
\cite{TL14} have found that star formation episodes can be suppressed, and 
the gas is kept pristine until the halo virial temperatures reach $\sim 8000~\K$ at $z\ga 30$\footnote{
The total gas mass is $\sim 7\times 10^5~\msun (T_{\rm vir}/8000~\K)^{3/2}[(1+z)/31]^{-3/2}$.
Thus, this fact requires a large fraction of the total mass to be converted 
into a single SMS with $M_\star >10^5~\msun$.}.
\cite{Hirano_Sci_2017} have found that a high gas accretion rate of $\dot{M} \ga 10^{-2}~\msunyr$ 
onto the center of a massive halo under strong BSM would be realized.
Recent cosmological simulations by \cite{Schauer_2017} suggest that 
in a rare patch of the Universe with strong BSM,
star formation is suppressed, even more efficiently than expected in \cite{TL14}.
Another simulation result by \cite{Hirano_2018} suggests that 
{\it dynamical effects due to frequent mergers of gaseous halos violently disturb the collapsing region
and further prevent star formation}.
As a result, star formation can be quenched until DM halos become as massive as
$\simeq 10^8~\msun$ ($T_{\rm vir}\ga 2\times 10^4~\K$ at $15<z<20$), 
which is $\ga 10$ times more massive than the minimum cooling mass in a region having strong BSM,
estimated by the previous work \citep{Greif_2011, Stacy_2011,Fialkov_2012,TL14}.
In a massive pristine DM halo, more efficient cooling can induce deeper penetration of cold accretion flows 
\citep[e.g.,][and see \S\ref{sec:shock}]{Fernandez_2014},
and form dense shocks where a massive seed BH would be initiated.

Massive pristine DM halos with $\simeq 10^8~\msun$ not only form a massive seed BH,
but also form bright galaxies with $\simeq 10^6~\msun$ dominated by PopIII stars.
In previous work, formation of PopIII clusters/galaxies in halos with $\sim 10^8~\msun$
has been discussed, considering strong ionizing radiation to stall gas collapse into halos
instead of strong BSM \citep{Johnson_2010,Yajima_2017,Visbal_2017}.
However, the number density of massive PopIII galaxies is expected to be 
$\sim 10^{-7}~\mpc^{-3}$ at $z\sim 7$ \citep{Visbal_2017}, which seems too rare to be 
observed by the {\it James Webb Space Telescope} ({\it JWST}).
In contrast, the BSM effect enhances the number density of PopIII galaxies at 
$z>10$ by several orders of magnitude because DM halos under streaming velocity 
with twice the root-mean-square value are not as rare as DM halos exposed to very bright
and synchronously forming ionizing neighbors.
This result strongly motivates us to explore PopIII galaxies with JWST.

This paper is organized as follows. 
In \S\ref{sec:sm}, we describe the method to study the suppression of star formation in massive 
DM halos in rare patches of the Universe with high streaming velocity and unusual merger histories with
frequent major merger events.
In \S\ref{sec:cold}, we discuss the necessary conditions required to form massive seed BHs in 
dense and hot shock regions produced at the interface between cold streams and a galactic disc,
and discuss the detectability of massive Pop III galaxies with JWST.
Finally, we discuss several caveats in \S\ref{sec:cav} and summarize our conclusions in \S\ref{sec:sum}.
Throughout this paper, we assume a CDM cosmology consistent with \cite{2014A&A...571A..16P}:
$\Omega_{\rm m}=0.32$, $\Omega_{\Lambda}=0.68$, $\Omega_{\rm b}=0.049$,
$h=0.67$, $\sigma _8=0.83$ and $n_{\rm s} =0.96$.

%%%%%%%%%
%	Section 2    %
%%%%%%%%%

\section{Violent mergers enhance the mass of pristine atomic-cooling halos
under high streaming velocities}
\label{sec:sm}

We consider a massive DM halo with a virial temperature of 
$T_{\rm vir}\ga 10^4~\K$ at high redshift.
The virial mass, the virial radius and the circular velocity of the halo are given by
\begin{equation}
M_{\rm vir}\simeq 1.7\times 10^7~T_{\rm vir,4}^{3/2}~\msun
\left(\frac{1+z}{16}\right)^{-3/2},
\end{equation}
\begin{equation}
R_{\rm vir}\simeq 350~h^{-1}~T_{\rm vir,4}^{1/2}~\pc
\left(\frac{1+z}{16}\right)^{-3/2},
\end{equation}
and 
\begin{align}
V_{\rm vir}=\sqrt{\frac{GM_{\rm vir}}{R_{\rm vir}}}
\simeq 12~T_{\rm vir,4}^{1/2}~{\rm km~s^{-1}}
\end{align}
\citep{Bryan_Norman_1998}.

\begin{figure}
\begin{center}
\includegraphics[width=83mm]{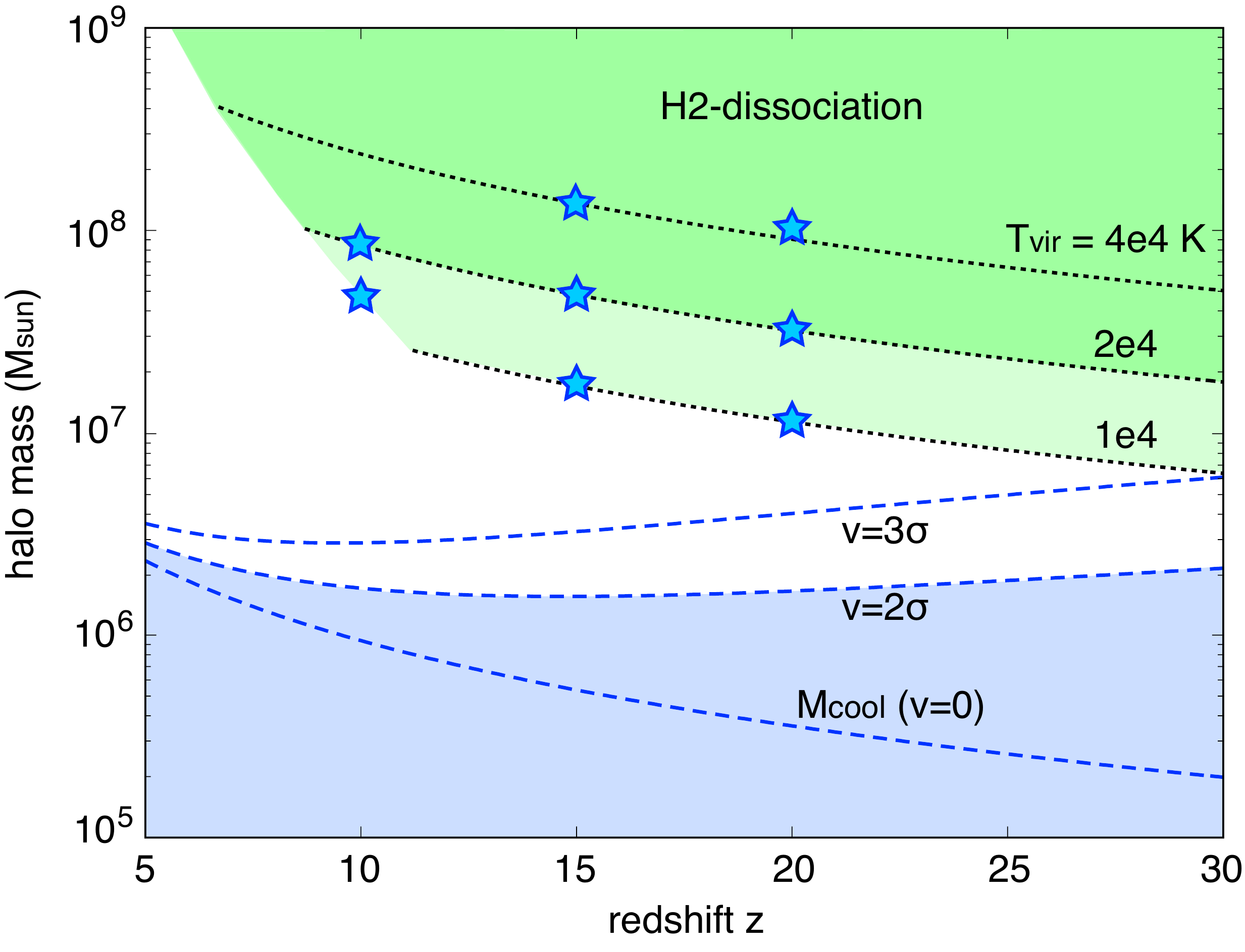}
\caption{Evolution of characteristic DM halo masses.
Dashed curves (blue) show the critical halo masses above which 
the gas can collapse into the halo, for three different streaming velocities $v_{\rm bsm}$.
In the green region, H$_2$ is collisionally dissociated if dense shocks form at the edge 
of the galactic disc (see Eq. \ref{eq:shock_cosm}).
Dotted curves indicate constant virial temperatures, the values of which are denoted by 
numbers in the figure.
The critical virial temperature required for deep penetration of cold streams 
is set to $T_{\rm vir}\geq 2\times 10^4~\K$ (see \S\ref{sec:BH}).
Stars show the masses and redshifts of DM halos we simulated 
with merger trees (see Table 1).
The gas between the blue and green regions can be bridged by halos with unusually frequent merging histories.
}
\label{fig:z_M}
\end{center}
\end{figure}

We here consider a rare patch of the Universe where BSM suppresses star formation episodes 
in the progenitor halos.
The streaming velocity $v_{\rm bsm}$ follows a Maxwell-Boltzmann distribution with the root-mean-square
speed of $\sigma_{\rm bsm}\sim 30~\kms$ at $z_{\rm rec}\simeq 1100$, and decays as 
$\tilde{v}_{\rm bsm}(z)=v_{\rm bsm} [(1+z)/(1+z_{\rm rec})]$ toward lower redshift \citep[e.g.,][]{TH_2010}.
Because of the BSM, star formation in mini-halos with $M_{\rm vir}\la 10^6~\msun$ is significantly delayed until $z<10$.
\cite{Fialkov_2012} fitted the critical circular velocity of DM halos above which the gas can collapse as 
\begin{equation}
v_{\rm cool}=\sqrt{v_0^2+[\alpha \tilde{v}_{\rm bsm}(z)]^2},
\label{eq:v_cool}
\end{equation}
where the two parameters are estimated as $(v_0,\alpha)\simeq (3.7~\kms, 4)$,
based on the results with cosmological simulations for PopIII formation in mini-halos
\citep{Greif_2011,Stacy_2011}.
Using Eq. (\ref{eq:v_cool}), the critical halo mass for cooling is given by
\begin{align}
M_{\rm cool}(v_{\rm bsm},z)&=1.9\times 10^6~\msun \left(\frac{1+z}{16}\right)^{-3/2}\nonumber\\
&\times \left[ 0.43 + \left(\frac{v_{\rm bsm}}{3\sigma_{\rm bsm}}\right)^2 \left( \frac{1+z}{16}\right)^2\right]^{3/2}.
\label{eq:M_cool}
\end{align}
Note that those simulations considered either weak BSM $v_{\rm bsm}\simeq \sigma_{\rm bsm}$ 
or an isolated DM halo (i.e., a small simulation box).
In Fig. \ref{fig:z_M}, we show the critical masses $M_{\rm cool}(v_{\rm bsm},z)$ 
for three different streaming velocities (blue dashed curves).

%%%%%%%%%
%	Fig. 1	%
%%%%%%%%%

Recently, \cite{Schauer_2017} have studied the effect of a high streaming velocity with
$v_{\rm bsm}=3\sigma_{\rm bsm}$ considering a large simulation box ($64~h^{-3}~\mpc^3$),
and found that the gas in DM halos less massive than $\simeq 2\times 10^7~\msun$ 
can neither collapse nor form stars.
Similarly, \cite{Hirano_2018} concluded that in a rare patch of the Universe with 
$v_{\rm bsm}\ga 2\sigma_{\rm bsm}$,
the first collapsing object (with gas) is as massive as $\ga 3\times 10^7~\msun$ at $15\la z\la 20$.
The virial temperature of the halo is $T_{\rm vir}\ga 2\times 10^4~\K$.
These critical masses found by their cosmological simulations are $\ga 10$ times higher than
$M_{\rm cool} (v_{\rm bsm}\ga 2\sigma_{\rm bsm},z)$, shown in Fig. \ref{fig:z_M}.

The existence of the mass gap implies that additional effects suppress gas collapse 
even if the halo mass exceeds the minimum cooling mass.
\cite{Hirano_2018} have pointed out that frequent merger events of DM halos containing gas 
violently disturb the central collapsing core in the main progenitor, as shown in their Figure 2. 
Motivated by this, we here discuss how likely such violet mergers can quench star formation further 
and close the mass gap.
For this purpose, we use the Monte Carlo merger trees to simulate the assembly history of DM halos of interest. 
The algorithm of merger trees is detailed in \cite{TL13}: the formation of DM halos is based on the ellipsoidal 
collapse model \citep{Sheth_Tormen_2002} under the extended Press-Schechter formalism. 
When compared with cosmological N-body simulations, the ellipsoidal model is better than the
spherical model in predicting the number of most massive halos for large look-back times.
The algorithm of the merger tree is the ``method B" in \cite{Zhang_2008}. 
In this algorithm, the time step $\Delta z$ for the merger tree can be chosen freely.
We choose $\Delta z=0.02$ in this paper.

We have selected several atomic-cooling DM halos, with masses and redshifts ($M_0$, $z_0$). 
For each merger tree of these halos, we keep track of all the progenitors with masses above $M_{\rm cool}$. 
We are looking for the {\it pristine trees} 
that satisfies the following criteria: none of the progenitors have gas collapse. 
This requires that every progenitor above $M_{\rm cool}$ has major mergers frequently enough 
that the redshift interval between two consecutive major mergers is shorter 
than the timescale of star formation, $\Delta z_{\rm m} < \Delta z_{\rm col}$
(see Eq. \ref{eq:tcol} below). 
We here define major mergers as ones with progenitor mass ratio of 
$f_{\rm m}\equiv M_1/(M_1+M_2)\geq 1/4$ (i.e., $M_1\geq M_2/3$),
where $M_1$ and $M_2(\geq M_1)$ are masses of DM halos that undergo mergers.
For example, suppose a major merger occurring at redshift $z_1$. 
Then the progenitors of this event should experience another prior major mergers within
[$z_1, z_1 + \Delta z_{\rm col}(z_1)$]\footnote{In principle, the interval should be 
[$z_1$, $z_2$], where $z_2$ is obtained from $z_2 = z_1 + \Delta z_{\rm col} (z_2)$. 
But in practice, this is a second-order effect and makes little difference to the results.}, 
otherwise this tree is counted as a {\it metal-polluted tree} due to star formation. 
We keep the simulation until all the progenitors are below $M_{\rm cool}$,
or until the tree becomes metal-polluted.
Eventually, this calculation gives us the probability of pristine trees, 
$\mathcal{P}_{Z=0}$, for the parent halo with $M_0$ and $z_0$.
For each of these halos, we have many realizations of its merger trees, 
$N_{\rm tree}\sim O(10^4-10^6)$, so that the statistical errors become smaller than 
$0.1~\mathcal{P}_{Z=0}$.

\begin{table}
\caption{Results of merger-tree simulations}
\begin{center}
  \begin{tabular}{ccc|c|c|c|c}
  \hline
$v_{\rm bsm}$ & $T_{\rm vir,4}$ & $M_{0,7}$ & $z_0$ & $\mathcal{P}_{Z=0}$ & 
   $\mathcal{N}~(\cmpc^{-3})$\\
 \hline
\hline
 $60$ &  $1.0$  & 1.13 & 20 & $0.44$ 				& $2.5\times 10^{-3}$\\
 $60$ & $1.0$  & 1.70 & 15 & $0.11$ 				& $2.7\times 10^{-3}$\\
 $60$ & $1.4$  & 4.70 & 10 & $4.4 \times 10^{-5}$  	& $3.2\times 10^{-6}$\\
\hline
 $60$ &  $2.0$  & 3.2 & 20 & $0.12$			 	& $2.3\times 10^{-4}$\\
 $60$ &  $2.0$  & 4.8 & 15 & $4.8 \times 10^{-3}$ 	& $5.8\times 10^{-5}$\\
 $60$ &  $2.0$  & 8.4 & 10 & $<5.0 \times 10^{-7}$ 	& $<1.8\times 10^{-8}$\\
\hline
 $60$ & $4.3$  & 10 & 20 & $4.9 \times 10^{-3}$ 	& $1.0\times 10^{-6}$\\
 $60$ & $4.0$  & 14  & 15 & $6.0 \times 10^{-6}$	 	& $1.0\times 10^{-8}$\\
 \hline
 $30$ &  $2.0$  & 4.8 & 15 & $4.0\times 10^{-6}$ 	& $2.6\times 10^{-6}$\\
 $90$ &  $2.0$  & 4.8 & 15 & $8.9 \times 10^{-2}$ 	& $8.0\times 10^{-7}$\\
 $60 \star$ &  $2.0$  & 4.8 & 15 & $5.7\times 10^{-5}$ 	& $7.0\times 10^{-7}$\\
 \hline
  \end{tabular}
  \end{center}
  {Column: (1) streaming velocity in unit of $\kms$ ($\sigma_{\rm bsm}=30~\kms$), 
  (2) virial temperature in unit of $10^4~\K$, 
  (3) DM halo mass in unit of $10^7~\msun$, 
  (4) redshift for DM halos we simulate, (5) the fraction of merger trees in which
  the gas is kept pristine due to violent major mergers, $\mathcal{P}_{Z=0}$, and (6) their number density, 
  $\mathcal{N}\equiv f_{v} \mathcal{P}_{Z=0} (dn_{\rm h}/d\ln M_{\rm vir})$.
  We adopt $C=1.5$ to characterized the star formation timescale (Eq. \ref{eq:tcol}), 
  calibrated based on simulations by \cite{Hirano_2018} (see text).
  In the last row ($\star$), we set $C=1$ to show the lowest probability.
  }
\end{table}

In order to obtain $\Delta z_{\rm col}$, we here estimate the timescale of star formation as
\begin{equation}
\Delta t_{\rm col}=C\sqrt{\frac{3\pi}{32G \bar{\rho}_{\rm m}(z)\Delta_{\rm vir}}},
\label{eq:tcol}
\end{equation}
where $\bar{\rho}_{\rm m}(z)$ is the mean mass density (DM+gas) at redshift $z$,
$\Delta_{\rm vir}\simeq 18\pi^2$ is the over-density relative to the cosmic mean value
at the collapse redshift.
We note that the expression of $t_{\rm ff}(\rho)=\sqrt{3\pi/(32G\rho)}$ is the free-fall timescale
of an initially static and uniform medium with a mean density $\rho$.
The actual collapse timescale of the cloud is longer than that for 
pressure-free collapse $t_{\rm ff}$.
The factor $C(>1)$ characterizes the delay of collapse due to 
the gas pressure gradient force and residual of BSM.
When an isothermal cloud in a hydrostatic equilibrium collapses, 
the collapse timescale becomes longer than the free-fall one and thus
$C\simeq 2-3$ \citep{Foster_1993,Tsuribe_1999,Aikawa_2005}.
Assuming that a major merger occurs at $z$, 
the redshift interval where gas collapse stalls is estimated as
\begin{align}
\frac{\Delta z_{\rm col}}{1+z} =\left[ 1-\left\{ 1 + 
\frac{\Delta t_{\rm col}(z)}{t_{\rm H}(z)}\right\}^{-2/3} \right] \equiv D,
\end{align}
where the ratio of $\Delta t_{\rm col}(z)$ 
to the Hubble timescale $t_{\rm H}(z)$ does not depend on redshift as
\begin{align}
\frac{\Delta t_{\rm col}(z)}{t_{\rm H}(z)}
=\frac{C}{4\sqrt{2}}.
\end{align}
We define the redshift when gas can collapse unless next major mergers occur 
as $\bar{z} \equiv z - \Delta z_{\rm col}$.
Thus, the redshift interval is rewritten as 
\begin{align}
\Delta z_{\rm col} &= \frac{D(1+\bar{z})}{1-D}
\simeq (1+\bar{z}) \times 
\begin{cases}
    0.12 & (C=1), \\
    0.17 &(C=1.5), \\
    0.22 & (C=2).
    \label{eq:dzcol}
  \end{cases}
\end{align}
In the merger-tree calculations, where we are looking at the history backward, 
we adopt Eq.(\ref{eq:dzcol}) setting $\bar{z}=z_1$.
Taking the results from cosmological simulations by \cite{Hirano_2018},
where two DM halos are studied for $v_{\rm bsm}\geq  2~\sigma_{\rm bsm}$, 
the delay-timescale of star formation is $\Delta z_{\rm col}\sim 2-4$ at 
$15 \la \bar{z}\la 23$, which allows us to adopt $C\simeq 1.5$ as our fiducial value,
calibrated by these simulations.
We note that in the simulations by \cite{Hirano_2018}, the density at the central core 
($\la 0.1~R_{\rm vir}$) before collapse is as high as $\ga 10^2~\cc$, where the free-fall timescale 
is much shorter than that we consider.
However, the dense core cannot be gravitationally unstable 
until gas at larger scales $\sim R_{\rm vir}$ begins to collapse.
Thus, the dense core can be disrupted by major mergers, and then 
takes longer to reform (i.e. the collapse time at the mean halo density).

\begin{figure}
   \begin{minipage}{1.0\hsize}
        \hspace{0mm}
	  \includegraphics[width=83mm]{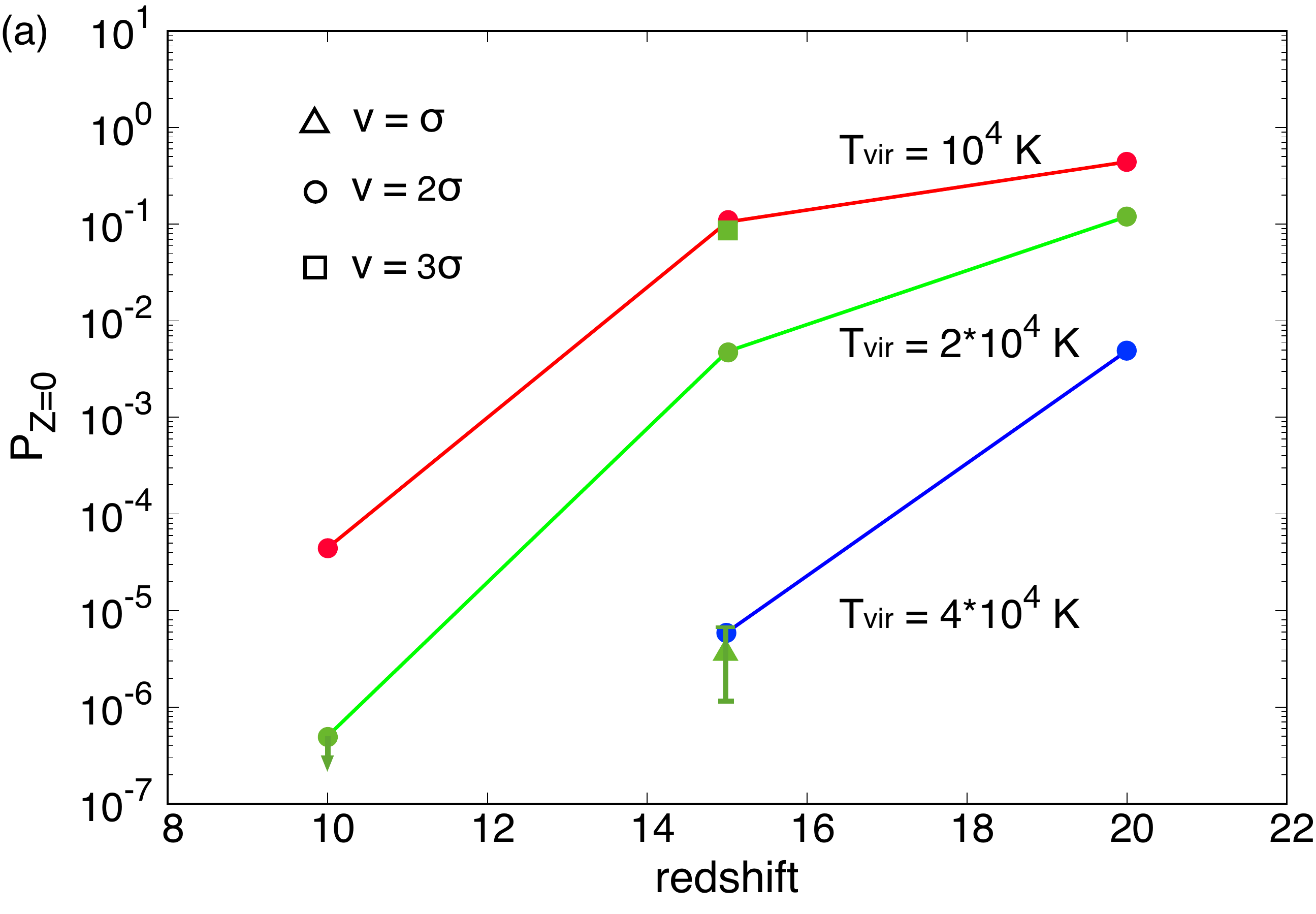}
      \end{minipage}    
   \begin{minipage}{1.0\hsize}
          \vspace{2mm}
        \hspace{0mm}
	  \includegraphics[width=83mm]{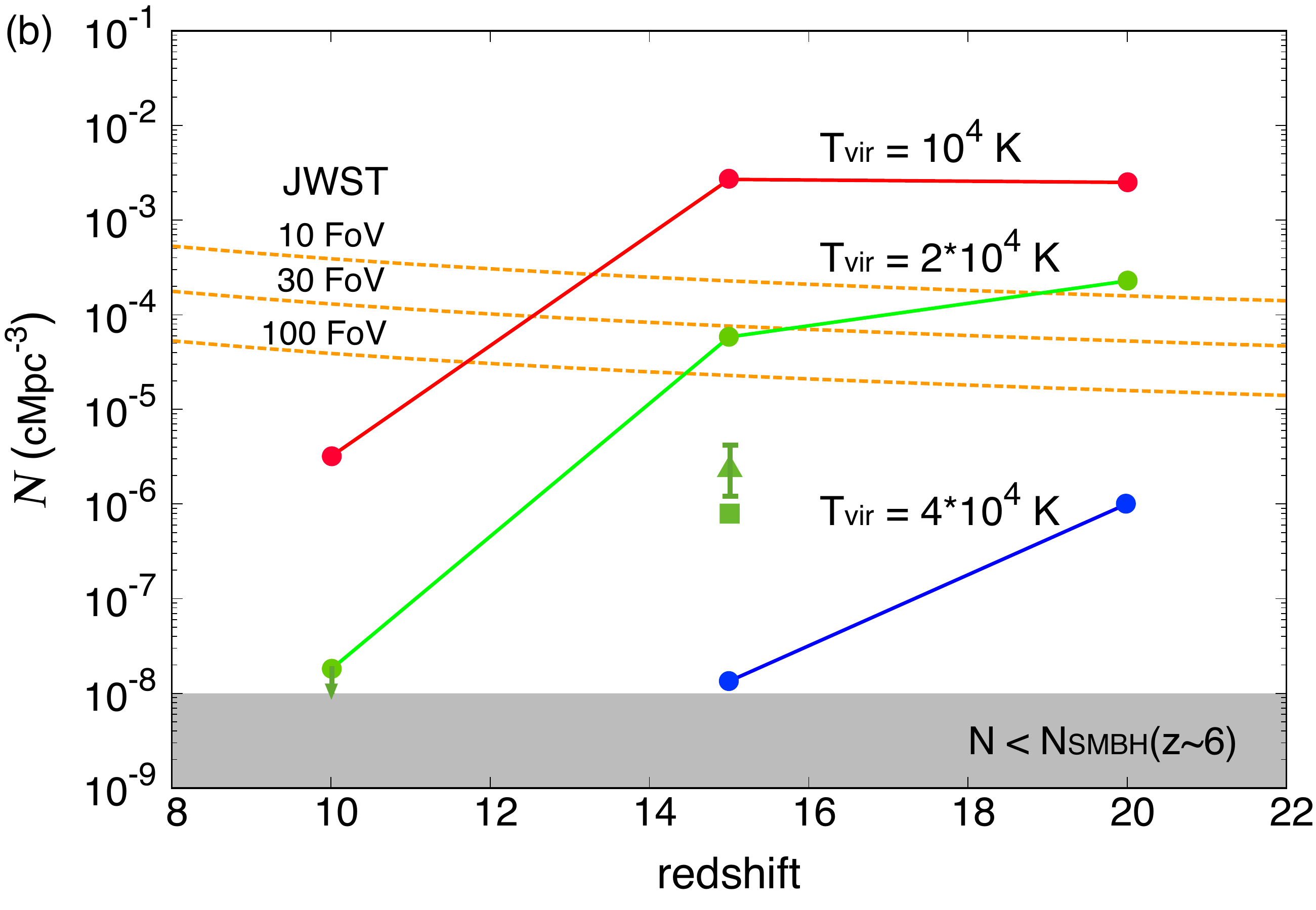}
             \end{minipage}
  \caption{(a) Redshift-evolution of the probability for a DM halo to remain pristine
  as a result of suppressing star formation due to strong BSM with $v_{\rm bsm} =2\sigma_{\rm bsm}$ (circle)
  and violent megers of gaseous halos caused by the BSM.
  Each line corresponds to the case with $T_{\rm vir}\simeq 10^4~\K$ (red), $2\times 10^4~\K$ (green),
  and $4\times 10^4~\K$ (blue).
  The probability decreases for more massive halos at lower redshift
  because the mass differences with the minimum cooling mass become larger 
  (see the green and blue regions in Fig.~\ref{fig:z_M}).
  The dependence on streaming velocities is shown for $T_{\rm vir}=2\times 10^4~\K$ at $z=15$:
  $v_{\rm bsm}/\sigma_{\rm bsm}=1$ (triangle) and $3$ (square). 
  For each of these halos, we have many realizations of its merger trees 
  so that the statistical errors become smaller than $0.1~\mathcal{P}_{Z=0}$ 
  (except for a case with $v_{\rm bsm}=\sigma_{\rm bsm}$).
  (b) The number density of massive pristine DM halos as a function of redshift.
  Orange curves show the minimum number densities required for a PopIII galaxy to be 
  detected by JWST, assuming a duty cycle of PopIII galaxies (see Eq. \ref{eq:pop3gal_n}),
  within three different fields of view $\Delta \Omega_{\rm obs}=10$, $30$ and $100$ FoV 
  from the top to bottom, where FoV $=9.7$ arcmin$^2$.
  In the gray region, the number density is lower than that of high-z SMBHs ($\sim 1-10~\gpc^{-3}$).
}
  \label{fig:ptree}
\end{figure}

In Fig. \ref{fig:ptree} and Table. 1, we summarize the results of the merger-tree simulations 
for our fiducial case ($C=1.5$).
For a given DM halo with $M_0$ and $z_0$,
we calculate the fraction of merger trees $\mathcal{P}_{Z=0}$ that keep the gas pristine 
due to violent major mergers even after the masses of the all progenitors exceed 
the minimum cooling mass $M_{\rm cool}$.
As shown in Fig. \ref{fig:ptree} (a), the fraction of pristine DM halos increases 
for lower $T_{\rm vir}$ or higher $z_0$ (i.e., lower $M_0$) because the mass gap between 
$M_{\rm cool}$ and $M_0$ becomes smaller (see the blue and green hatched regions in Fig. \ref{fig:z_M}).
For higher $T_{\rm vir}$ or lower $z_0$ (i.e., higher $M_0$), the probability drops sharply
because the mass gap becomes bigger.
Combined with the mass function of DM halos \citep{Sheth_Tormen_2002}, 
we can estimate the number density of DM halos containing pristine gas as
\begin{align}
\mathcal{N} =  f_{2 \sigma} \mathcal{P}_{Z=0} 
\frac{dn_{\rm h}}{d\ln M_{\rm vir}},
\label{eq:n_pac}
\end{align}
where the fraction of the Universe with streaming velocities of $v_{\rm bsm}\geq 2\sigma _{\rm bsm}$ 
is $f_{2\sigma} \simeq 8\times 10^{-3}$.
Note that the coherence length of the velocity field is $\sim 10$ Mpc \citep{TH_2010},
which is much larger than the typical separation between atomic-cooling halos of interest.
Thus, the root-mean-square value of the streaming velocity can represent those at the locations of halos.
Intriguingly, the number densities of massive pristine DM halos at $15\la z \la 20$ are
$\mathcal{N}\simeq 10^{-5}-10^{-4}~\mpc^{-3}$ for $T_{\rm vir}\sim 2\times 10^4~\K$, and 
$\mathcal{N}\simeq 10^{-8}-10^{-6}~\mpc^{-3}$ for $T_{\rm vir}\sim 4\times 10^4~\K$.

\begin{table}
\caption{Same as Table 1 but for different criteria for characterizing major mergers of DM halos}
\begin{center}
  \begin{tabular}{cccc}
  \hline
  $f_{\rm cr}$ & mass ratio & $\mathcal{P}_{Z=0}$ & 
   $\mathcal{N}~(\cmpc^{-3})$\\
 \hline
\hline
 $0.15$ & $1:5.67$ & $5.4\times 10^{-2}$ & $6.5\times 10^{-4}$\\
 $0.2$ & $1:4$      & $1.6\times 10^{-2}$ & $1.9\times 10^{-4}$\\
 $0.25$ & $1:3$      & $4.8\times 10^{-3}$ & $5.8\times 10^{-5}$\\
 $0.3  $ & $1:2.33$ & $1.5\times 10^{-3}$ & $1.8\times 10^{-5}$\\
\hline
  \end{tabular}
  \end{center}
  {Major mergers are defined as ones with progenitor mass fraction of 
  $f_{\rm m}\geq f_{\rm cr}$, where $f_{\rm m}\equiv M_1/(M_1+M_2)$ and $M_1\leq M_2$. 
  We show the results for $T_{\rm vir}=2\times 10^4~\K$, $M_{0}=4.8\times 10^7~\msun$, 
  $z_0=15$ and $v_{\rm bsm}=60~\kms$.
  The fraction of pristine trees is fitted by $\mathcal{P}_{Z=0}=\exp(-Af_{\rm cr}+B)$, where 
  $A=23.85$ and $B=0.64$.
  }
\end{table}

We further investigate the dependence of our results on the streaming velocity,
focusing on DM halos with $M_0=4.7\times 10^7~\msun$ and $z_0=15$ ($T_{\rm vir}=2\times 10^4~\K$).
As shown in Figure \ref{fig:ptree}(a) and Table 1, The probability of pristine halos increases (decrease) 
for higher (lower) $v_{\rm bsm}$ because the mass gap between $M_{\rm cool}$ and $M_0$ becomes smaller (larger).
we obtain $P_{Z=0}\simeq (4.0 \pm 2.8) \times 10^{-6}$ for $v_{\rm bsm}=\sigma_{\rm bsm}$ and 
$P_{Z=0}=8.9\times 10^{-2}$ for $v_{\rm bsm}=3\sigma_{\rm bsm}$.
Combined with the volume fraction of the Universe with the streaming velocities
($f_{1\sigma} \simeq 0.4$ and $f_{3\sigma} \simeq 5.9\times 10^{-6}$),
the number density of DM halos of interest is estimated as 
$\mathcal{N}\simeq (2.6 \pm 1.5) \times 10^{-6}~\mpc^{-3}$ for $v_{\rm bsm}=\sigma_{\rm bsm}$ and
$\mathcal{N}\simeq 8.0 \times 10^{-7}~\mpc^{-3}$ for $v_{\rm bsm}=3\sigma_{\rm bsm}$, respectively
(see Figure \ref{fig:ptree}b).
As a result, we find that the number density can be maximized at $v_{\rm bsm}\simeq 2\sigma_{\rm bsm}$.
Therefore, we adopt $v_{\rm bsm}= 2\sigma_{\rm bsm}$ as our fiducial case in the following sections.

%%%%%%%%%
%	Fig. 2	%
%%%%%%%%%

Compared to previous work with cosmological simulations, we can justify our choice of $C(=1.5)$.
\cite{Schauer_2017} have found 36 DM halos with $M_{\rm vir}\ga 3\times 10^7~\msun$ at $z=15$,
where gas has just collapsed, in a rare patch of the Universe with $v_{\rm bsm}=3\sigma_{\rm bsm}$.
Since the volume of their simulation is $(4h^{-1}~\cmpc)^3$, the number density of massive DM halos 
containing pristine gas is $\sim 1.1\times 10^{-6}~\cmpc^{-3}$ at $z=15$,
which is consistent with our result for $C=1.5$ ($\mathcal{N}\simeq 8.0 \times 10^{-7}~\mpc^{-3}$) 
or rather implies $C\ga 1.5$.

It is also worth estimating the lower bound of the probability by assuming $C=1$, 
which corresponds to the pressure-free collapse.
For $v_{\rm bsm}\simeq 2\sigma_{\rm bsm}$, we obtain $P_{Z=0}\sim 5.7\times 10^{-5}$
for DM halos with $M_0=4.7\times 10^7~\msun$ at $z_0=15$.
Then, this probability $\sim 80$ times smaller than that for $C=1.5$, and the number density of DM 
halos of interest is $\sim 7.0\times 10^{-7}~\cmpc^{-3}$.

The probability for a DM halo to remain pristine depends on the critical mass fraction $f_{\rm cr}$, 
above which the merger events can suppress star formation temporarily 
($f_{\rm cr}=1/4$ is our fiducial value).
The results are summarized in Table 2.
The fraction of DM halos where the gas remains pristine decreases with $f_{\rm cr}$ and is fitted by 
$\mathcal{P}_{Z=0}=\exp(-Af_{\rm cr}+B)$, where $A=23.85$ and $B=0.64$.
For $f_{\rm cr}\la 0.3$, the fraction is sufficiently high for the detection of such objects by JWST (see \S\ref{sec:gal}).
In order to quantify the critical mass ratio for quenching star formation due to halo mergers, hydrodynamical simulations are necessary, which we leave to future work.

%%%%%%%%%
%	Section 3	%
%%%%%%%%%

\section{Formation of massive seed BHs and Population III galaxies}
\label{sec:cold}

\subsection{Cold accretion streams}

In a rare region of the high-$z$ Universe, where strong BSM significantly suppresses star formation
as discussed in \S\ref{sec:sm}, the gas can be kept primordial even in a massive DM halo with 
$T_{\rm vir}\ga 10^4~\K$.
The inflow rate of the pristine gas from larger scales is estimated as
\begin{equation}
\dot{M}_{\rm gas}\simeq f_{\rm b}\frac{V_{\rm vir}^3}{G}
\simeq 6.6\times 10^{-2}~T_{\rm vir,4}^{3/2}~\msunyr,
\label{eq:Mdot_halo}
\end{equation}
\citep[e.g.,][]{White_Frenk_1991, Mayer_2015},
where the baryon fraction, $f_{\rm b}\equiv \Omega_{\rm b}/\Omega_{\rm m}$ is set to $0.16$.
A significant fraction of the gas can accrete onto the galactic center through cold streams.
We define the gas accretion rate along one stream as
$\dot{M}_{\rm s}=\epsilon _{\rm s} \dot{M}_{\rm gas}$, where the fraction of $\epsilon_{\rm s}(<1)$ is 
given by $1/$(number of streams)
and is set to $\epsilon_{\rm s}\simeq 0.3$ as our fiducial value.

%%%%%%%%%
%	Fig. 3	%
%%%%%%%%%
\begin{figure}
\begin{center}
\includegraphics[width=83mm]{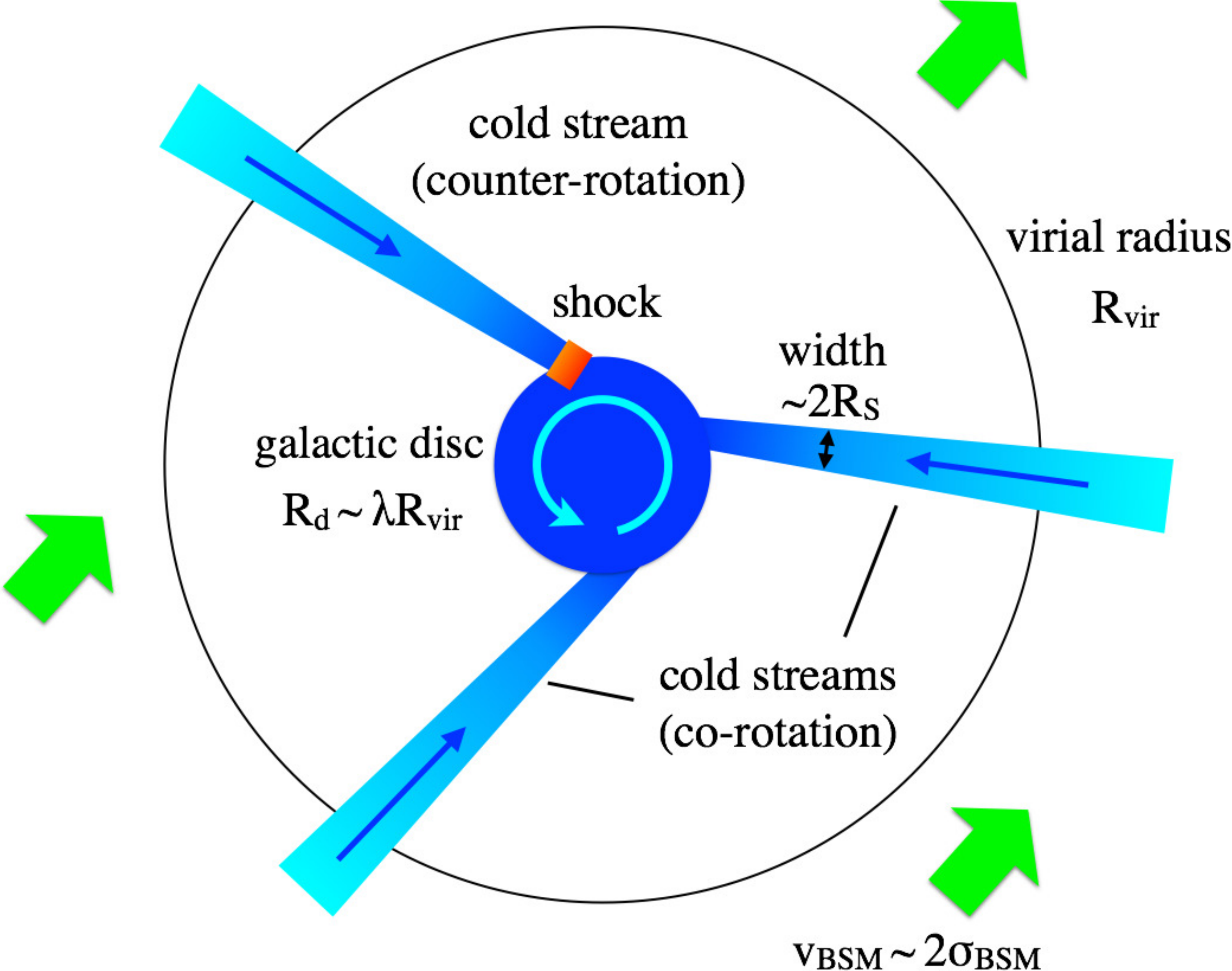}
\vspace{3mm}
\caption{A schematic picture of a massive DM halo fed by cold streams along large-scale DM filaments
in a rare patch of the Universe with a high streaming velocity of $v_{\rm bsm}=2\sigma_{\rm bsm}$.
The cold streams with a half width of $R_{\rm s}$ penetrate inside the virial radius of $R_{\rm vir}$ 
to the vicinity of the central galactic disc with a radius of $R_{\rm d}\simeq \lambda R_{\rm vir}$. 
Cold streams co-rotating with the host galaxy feed the central disc smoothly and 
lead to active star formation of ordinary PopIII stars.
On other hand, the counter-rotating cold stream collides with the disc and/or other streams,
forming strong shocks.
If the shocked gas is hot and dense enough to dissociate H$_2$ 
(see Eqs. \ref{eq:shock} and \ref{eq:shock_cosm}),
a massive self-gravitating cloud forms at the interface and collapse directly into a massive seed BH.}
\label{fig:diagram}
\end{center}
\end{figure}

According to cosmological simulations of galaxy formation, the typical stream width inside the halo 
is a few percent of the virial radius \citep{Keres_2005,Ocvirk_2008,Dekel_2009,Ceverino_2010}.
The cold accretion flow becomes denser at smaller radii because of contraction in the DM gravitational potential.
As a result of this, the geometry of the flow seems a conical one rather than cylindrical one.
Assuming a conical stream with an opining angle of $\theta_{\rm s}$, 
a half of the stream width is given by
$R_{\rm s}\simeq r\theta_{\rm s}/2 =0.1r(\theta_{\rm s}/0.2)$,
where $r$ is the distance from the center of the halo.
Such narrow cold streams with supersonic velocities penetrate well inside 
the virial radius, supplying gas and angular momentum towards the nuclear galactic disc.
In the interface region between the streams and disc, the gas dynamics is governed by 
a combination of hydrodynamical, thermal and gravitational instabilities, shocks in the streams, 
and interactions of streams with other streams and/or the inner disc \citep[e.g.,][]{Dekel_Sari_2009,Ceverino_2010}.
\cite{Danovich_2015} have investigated the nature of cold-stream penetration inside 
massive galaxies with $M_{\rm vir}\simeq 10^{11}-10^{12}~\msun$ at $1<z<4$.
They found that while most cold streams coherently join the rotationally supported central disc, 
a significant fraction ($\sim 30~\%$) of the cold gas has a counter-rotation component with 
respect to the total angular momentum.
In other words, for an atomic cooling halo fed by three cold streams (i.e., $\epsilon_{\rm s}\simeq 0.3$),
one of them settles onto the central disc with experiencing strong shocks (see \S\ref{sec:BH}),
but the other two smoothly supply gas to the disc (see \S\ref{sec:gal}).

The size of the galactic disc is characterized by the DM spin parameter of $\lambda_{\rm dm}$ as 
\begin{align}
R_{\rm d}=\lambda_{\rm dm} R_{\rm vir},
\label{eq:disc_r}
\end{align}
\citep{Fall_Efstathiou_1980,Mo_1998}.
Here, we assume that the angular momentum of cold streams is the same as that of DM halos.
\cite{Danovich_2015} have pointed out that for massive galaxies at $1<z<4$, 
the elongated streams gain angular momentum by tidal torques more than the DM halos,
in fact, $\lambda_{\rm gas}\sim (2-3)\lambda_{\rm dm}$ at $r\ga R_{\rm vir}$.
Thus, cold streams with a higher angular momentum dissipate at a relatively large radius of
$\sim 0.3(\lambda_{\rm dm}/0.15)~R_{\rm vir}$.
However, this effect becomes less important for high-redshift galaxies (see their figure 7),
which is consistent with the spin properties of mini-halos at $10<z<25$, where the mean spin parameters 
are estimated as $\lambda_{\rm gas}=0.0498$ and $\lambda_{\rm dm}=0.0495$ \citep{Hirano_2014}.
Therefore, we assume $\lambda_{\rm gas}\simeq \lambda_{\rm dm}~(\equiv \lambda)$ and adopt 
$\lambda=0.05$ as our fiducial model.

The inflow velocity of the cold stream does not follow the corresponding free-fall profiles.
Instead, the velocity can be approximated as $v_{\rm in}\simeq V_{\rm vir}$, 
independent of the distance from the center of the halo \citep{Goerdt_Ceverino_2015}.
In fact, several cosmological simulations for atomic-cooling halos suggest
that the inflow velocity is $\sim 15-20~\kms$ inside the virial radii
\citep{Greif_2008,Wise_2008} and near the edge of the central disc \citep{Schauer_2017}.
The velocity is as high as the circular velocity, i.e., 
$v_{\rm in} = \epsilon_{\rm v}V_{\rm vir}$.
We set $\epsilon_{\rm v}=1.5$ as our fiducial value.
Assuming steady accretion, we can estimate the gas density of 
a cone-like cold stream at the outer edge of the disc ($r=R_{\rm d}$) as 
\begin{align}
\rho  &=\frac{4\dot{M}_{\rm s}}{\pi v_{\rm in}(\theta_{\rm s}R_{\rm d})^2},\nonumber\\
&\simeq 3.7\times 10^{-21}~
\hat{\lambda}^{-2}~
\hat{\theta}_{\rm s}^{-2}
\left(\frac{1+z}{16}\right)^{3}
~{\rm g~\cc},
\label{eq:rho_disc}
\end{align}
where $\hat{\lambda}=\lambda/0.05$ and $\hat{\theta}_{\rm s}=\theta_{\rm s}/0.2$.

The gas density inside the central gaseous disc is high enough to form H$_2$ molecules
in the absence of strong shocks.
A large amount of pristine gas with $\sim {\rm a~few}\times 10^6~\msun$ 
(see Eq. \ref{eq:M_disc_pop}) forms ordinary PopIII stars with 
$M_\star \sim 10-100~\msun$ and evolve to a massive PopIII galaxy.
On the other hand, if cold streams with counter-rotation components dissipate due to shocks, 
H$_2$ is dissociated efficiently depending on the post-shock density and form an SMS with 
$\sim 10^5~\msun$, which eventually collapses into a massive seed BH.
In the following sections of \S\ref{sec:BH} and \ref{sec:gal}, 
we discuss the subsequent evolution pathways in the two distinct cases.

\subsection{Massive seed BH formation via colliding cold streams with the central disc}
\label{sec:BH}

A cold stream with a counter-rotation component with respect to the total angular momentum
is likely to collide with the outer edge of the galactic gaseous disc and/or other streams near the disc.
The temperature of the cold stream is as low as $T\la 8000~\K$ due to Ly$\alpha$ cooling and H$_2$ cooling.
Since the sound speed is estimated as $c_{\rm s}\simeq 2.6~{\rm km~s^{-1}}(T/10^3~\K)^{1/2}$,
the Mach number is as high as $v_{\rm in}/c_{\rm s}\sim 4.6\epsilon_{\rm v}(T_{\rm vir}/10T)^{1/2}$.
In such a strong shock limit, the gas density is enhanced by a factor of four, i.e., 
$\rho_{\rm post}\simeq 4\rho$ behind the shock front at $r\simeq R_{\rm d}$,
\begin{align}
n_{\rm post} &\simeq  1.4\times 10^4~
\hat{\lambda}^{-2}
\hat{\theta}_{\rm s}^{-2}
\left(\frac{1+z}{16}\right)^{3}
~\cc,
\end{align}
where the mean molecular weight of the postshock gas is set to $0.64$
(fully ionized hydrogen and neutral helium).
The post-shock temperature is estimated as 
\begin{align}
T_{\rm post}\simeq \frac{3\mu m_{\rm p}}{16 k_{\rm B}}v_{\rm in}^2
\simeq 8.4\times 10^3~T_{\rm vir,4}~\K.
\label{eq:shock}
\end{align}

We describe the necessary conditions for 
suppressing H$_2$ formation via collisional dissociation
in dense and hot shocked gas.
\cite{IO12} have studied thermal evolution of the shocked gas including chemical reactions 
among primordial species, and found that once the shocked gas enters a dense and hot region 
satisfying 
\begin{align}
T_{\rm post}>5.2\times 10^3~\K
\left(\frac{n_{\rm post}}{10^4~\cc}\right)^{-1},
\label{eq:post1}
\end{align}
the shocked gas never cools down via H$_2$ cooling because of H$_2$ collisional dissociation.
This condition can be rewritten in terms of the DM halo properties as
\begin{align}
T_{\rm vir,4} & >  0.44 ~
\hat{\lambda}^{2}
\hat{\theta}_{\rm s}^{2}
\left(\frac{1+z}{16}\right)^{-3}.
\end{align}
In addition, we require more massive haloes (i.e., higher virial temperature) 
to realize deep penetration of cold streams due to efficient radiative cooling.
Although the critical virial temperature is still uncertain, we conservatively assume 
the critical value to be $T_{\rm vir}= 2\times 10^4~\K$, making cooling timescale shorter.
Therefore, the necessary conditions for SMS formation are given by
\begin{align}
T_{\rm vir,4}\ga {\rm max}\left[ 
0.44~\hat{\lambda}^{2}\hat{\theta}_{\rm s}^{2}
\left(\frac{1+z}{16}\right)^{-3}
, ~2.0 \right],
\label{eq:shock_cosm}
\end{align}
which is shown in Fig. \ref{fig:z_M} (the hatched green region).
We note that the boundary of the H$_2$-dissociation dominant region is 
almost independent of the electron ($x_{\rm e}$) and H$_2$ fraction ($x_{\rm H_2}$)
in the pre-shock gas, as long as  $x_{\rm e}\ga 10^{-4}$ and $x_{\rm H_2}\la 10^{-3}$
(see Figure 2 in \citealt{IO12}).
However, the location of the boundary shifts to higher densities due to metal-line cooling
for $Z\ga 10^{-3}~\zsun$ (see Figure 3 in \citealt{IO12}).

Behind the shock front, the gas never cools below $T< 8000~\K$ because of the lack of H$_2$.
The shocked gas is accumulated and contract isobarically with the same ram pressure 
from the shock front, where $P_{\rm ext}\simeq \rho v_{\rm in}^2$.
An isothermal cloud embedded in an external pressure $P_{\rm ext}$ cannot exist in stable 
equilibrium if the mass of the cloud exceeds a critical value \citep[e.g.,][]{Ebert_1955,Bonnor_1956}:
\begin{align}
M_{\rm BE} 
&\simeq 1.05 \frac{\theta_{\rm s}\lambda}{\sqrt{\epsilon_{\rm s}\epsilon_{\rm v}f_{\rm b}}}
\left(\frac{c_{\rm s}}{V_{\rm vir}}\right)^{4} M_{\rm vir},\nonumber\\
&\simeq 9.6\times 10^{4}~
\hat{\lambda}~\hat{\theta}_{\rm s}~
T_{\rm vir,4}^{-1/2} ~
\left(\frac{1+z}{16}\right)^{-3/2}
~\msun.
\label{eq:MBE}
\end{align}
For our fiducial case, the shocked gas layer becomes massive enough to be 
unstable against its self-gravity in a timescale of $M_{\rm BE}/\dot{M}_{\rm s}\sim 
1.1~{\rm Myr}~(T_{\rm vir,4}/2)^{-2}$.
Since this timescale is sufficiently shorter than typical lifetime of massive stars, 
the SMS forming cloud would not be enriched by metals produced by supernovae,
which would happen in the central disc at $t\ga 4-10$ Myr (see discussions in \S\ref{sec:gal}).
After the onset of the gravitational instability, the gas cloud collapse in a runaway fashion
\citep[e.g.,][]{Larson_1969} and form a single protostar without major episodes of fragmentation
\citep[e.g.,][]{Regan_2014,IOT14,Becerra_2015,Latif_2016}.
The protostar grows via rapid accretion from the massive clump at a rate of 
$\dot{M}_\star \simeq 20~c_{\rm s}^3/G\simeq 2~\msunyr$ \citep[e.g.,][]{IOT14}.

\subsection{Massive PopIII galaxies formation via rapid feeding of pristine gas}
\label{sec:gal}

Next, let us consider the properties of a galaxy formed by metal-free gas
at the center of the most massive DM halo with $M_{\rm vir} \ga 5\times 10^7~\msun$ 
($T_{\rm vir}\ga 2\times 10^4~\K$ at $z\ga 15$).
Assuming that the mean gas density at $r\la R_{\rm d}$ is given by Eq. (\ref{eq:rho_disc}),
the disc mass is estimated as 
\begin{align}
M_{\rm d} &\simeq 0.085~\hat{\lambda}\hat{\theta}_{\rm s}
\sqrt{\frac{T}{T_{\rm vir}}}~M_{\rm vir}.
\label{eq:M_disc_pop}
\end{align}
For $T=8000~\K$ and $T_{\rm vir}=2\times 10^4~\K$, we obtain
$M_{\rm d}\simeq 0.054~\hat{\lambda}M_{\rm vir}
\simeq 0.34\hat{\lambda}(\Omega_{\rm b}/\Omega_{\rm m})~M_{\rm vir}$.
Therefore, we find that $M_{\rm d}\simeq 2.7\times 10^6\hat{\lambda}~\msun$ of gas 
is concentrated to the nuclear disc.
Note that the disc mass is consistent with that of a disc formed in a massive halo with 
$M_{\rm vir}\simeq 10^8~\msun$ at $z\simeq 15$ estimated with cosmological simulations 
without radiative feedback \citep{Pawlik_2011}.
The gas-rich disc becomes unstable due to the self-gravity, exciting bar modes to transport 
angular momentum and/or fragmenting to clumps.
The instability is characterized in terms of the Toomre's parameter $Q$, 
evaluated at $r=R_{\rm d}$ as
\begin{align}
Q=\frac{c_{\rm s}\kappa}{\pi G\Sigma_{\rm d}}
\simeq 0.63~\hat{\theta}_{\rm s}^{1/2}\left(\frac{T}{T_{\rm vir}}\right)^{1/4},
\end{align}
where $\Sigma_{\rm d}$ is the surface density of the disc,
and $\kappa$ is the epicyclic frequency, which is related to the orbital frequency $\Omega$
as $\kappa^2 = rd{\Omega^2}/dr +4\Omega^2$.
We here adopt $\kappa=\Omega$ for Keplerian orbits.
Note that for $T=8000~\K$ and $T_{\rm vir}=2\times 10^4~\K$,
we obtain $Q\simeq 0.5~(<1)$, for which fragmentation and star formation are triggered.
The gas density of each fragment is as high as $n\sim 10^3~\cc$ and thus the free-fall timescale of 
$\sim 1$ Myr is significantly shorter than the lifetime of $\sim 4-10$ Myr for massive PopIII
stars with $M_\star \simeq 10-50~\msun$ \citep[e.g., ][]{Marigo_2001}.
In the following, we consider the early stage of star formation within $\sim 3$ Myr
because (1) supernova feedback can be neglected and (2) the detectability of younger PopIII 
galaxies by JWST is higher.

The absence of supernova feedback in the early stage ($t\la 3$ Myr) enables 
a high star formation efficiency $\epsilon_\star \equiv M_{\star ,\rm tot}/M_{\rm d}$,
where $M_{\star ,\rm tot}$ is the total mass of PopIII stars formed in the disc.
For massive PopIII stars, ionizing radiation emitted from accreting protostars heats the ambient gas
and photo-evaporates their parent cloud \citep{McKee_Tan_2008,Hosokawa_2011}.
\cite{Hirano_2014} suggest that the properties of parent clouds, namely accretion rates onto protostars, 
determine the star formation efficiency and the initial mass function (IMF) of stars.
In fact, the efficiency is estimated as 
$\epsilon_\star \simeq 0.3$ for $10 \la M_\star /\msun \la 10^2$ (IMFa), and
$\epsilon_\star \simeq 0.5$ for $10^2 \la M_\star /\msun \la 10^3$ (IMFb).
Thus, the total stellar mass 
is estimated as
\begin{equation}
M_{\star,\rm tot}\simeq 
  \begin{cases}
    8\times 10^5\hat{\lambda}~\msun & {\rm (IMFa)},\\
    1.4\times 10^6\hat{\lambda}~\msun & {\rm (IMFb)}.
  \end{cases}
\end{equation}
We note that the star-formation efficiencies we assumed would be an upper limit
because ionizing radiation from newly-born stars could affect nearby star forming clouds
and reduce the total star formation efficiency in the galaxy.
With a spectral synthesis model for PopIII stars, \cite{Zackrisson_2011} discussed the detectability of 
PopIII galaxies through broadband imaging at $10~\sigma$ after a 100 hr exposure with JWST. 
In the early stage of a star-burst episode within $\sim 3$ Myr, 
the minimum mass of PopIII galaxies required for the detection is estimated as
\begin{equation}
M_{\star,\rm tot}^{\rm obs}\simeq 
  \begin{cases}
    (4-10)\times 10^5~\msun & {\rm (IMFa)},\\ 
    (2-3)\times 10^5~\msun & {\rm (IMFb)},
  \end{cases}
\end{equation}
where the lower (higher) value corresponds to that at $z=10~(15)$.
The exposure time for the detection is $t_{\rm exp}\simeq (6.3-40)\times (\lambda/0.1)^{-2}$ hr for IMFa, 
but is only $t_{\rm exp}\simeq (0.5-1.1)\times (\lambda/0.1)^{-2}$ hr for IMFb.
Although the exposure time is shorter for higher spin,
the free-fall timescale in the disc is comparable or longer than the lifetime of massive PopIII stars.
Therefore, our assumption neglecting the effects of supernova explosions breaks down for
$\lambda \ga 0.2$.
In addition, the conditions for massive seed BHs becomes more severe for $\lambda \ga 0.2$
because the post-shock density is not high enough to dissociate H$_2$.
Overall, we find that $\lambda =0.1$ and very top-heavy IMF (IMFb) are 
an interesting combination in terms of massive seed BH formation and
the detectability of PopIII galaxies with JWST.

We aim to detect massive PopIII galaxies at $10\la z \la 15$.
For this redshift range, Ly$\alpha$ and He$_{\rm II}~1640~{\rm \AA}$ emission 
can be observed by NIRCam of JWST 
\citep[e.g.,][]{Tumlinson_Shull_2000,Oh_Haiman_Rees_2001}.
As discussed in \cite{Zackrisson_2011}, we could identify young PopIII galaxies with 
the ages of $t_{\rm burst}\simeq 3$ Myr, using the JWST color criteria (see also \citealt{Inoue_2011}).
Thus, considering the duty cycle of PopIII star bursts, we can estimate the effective 
comoving volume where PopIII galaxies are detected as 
\begin{align}
\Delta V_{\rm obs}(z) &= 
\Delta \Omega_{\rm obs}
~\frac{dV_{\rm c}}{dz} 
~\frac{dz}{dt_{\rm r}} 
~t_{\rm burst},\label{eq:pop3gal_n}\\
& \simeq 4.4\times 10^3~\cmpc^{3}\left(\frac{t_{\rm burst}}{3~{\rm Myr}}\right)
\left(\frac{1+z}{16}\right)^{\gamma}
\left(\frac{\Delta \Omega_{\rm obs}}{{\rm 10~FoV}}\right),\nonumber
\end{align}
where $V_{\rm c}$ is the comoving volume, $t_{\rm r}$ is the time in a source's cosmic rest frame, 
$\Delta \Omega_{\rm obs}$ is the observed field of view (one FoV $\simeq 9.7$ arcmin$^2$), 
and $\gamma \simeq 1.406$ at $10\leq z\leq 20$.
In Fig. \ref{fig:ptree} (b), we show the number density of PopIII galaxies required for the detection 
in a given total FoVs is $\simeq 1/\Delta V_{\rm obs}(z) \simeq 2.3\times 10^{-4}~\cmpc^{-3}
[{(1+z)}/16]^{-\gamma}(\Delta \Omega_{\rm obs}/10~{\rm FoV})^{-1}$.
The threshold value is comparable to the number density of PopIII galaxies 
($n\sim 10^{-4}~\cmpc^{-3}$) with $T_{\rm vir}\simeq 2\times 10^4~\K$.
In order to detect one PopIII galaxy by JWST, we require $\simeq 1$ hr observations in 
different $\simeq 30$ areas, i.e., the total field of view $\sim 300$ arcmin$^2 \sim 0.1$ deg$^2$.
This total area corresponds to $\la 3~\%$ of Ultra deep field of Subaru Hyper Suprime Cam
\citep{Subaru_survey_2018}\footnote{https://hsc-release.mtk.nao.ac.jp/doc/}.

%%%%%%%%%
%	Section 4	%
%%%%%%%%%

\section{Discussion}
\label{sec:cav}

\subsection{Implications of subsequent BH evolution}

Once a seed BH with a mass of $M_\bullet \sim 10^5\msun$ forms
in a massive DM halo with $M_{\rm vir}\sim 10^8~\msun$,
subsequent growth via gas accretion and/or BH mergers is required to reach
$M_\bullet \sim 10^9~\msun$ by $z\simeq 6-7$ \citep[e.g.,][]{TH09,Valiante_2016,Pezzulli_2016}. 
Assuming the Eddington-limited accretion, the BH growth timescale from 
$10^5$ to $10^9~\msun$ is estimated as $\simeq 0.46$ Gyr, 
where the radiative efficiency is set to a constant value of $\eta = 0.1$. 
The growth time is marginally shorter than the cosmic time from $z = 20$ to $7$ ($\simeq 0.57$ Gyr)
\citep{HaimanLoeb01}.

\cite{Valiante_2016} have investigated the relative role of light seeds (i.e., PopIII remnant BHs) 
and heavy (massive) seeds as BH progenitors of the first SMBHs, conducting semi-analytical calculations.
They concluded that, as long as gas accretion is assumed to be Eddington-limited, 
heavy BH seeds at $z\ga 15$ are required to form SMBHs with $\ga 10^9~\msun$ by $z \simeq 6$.
Moreover, in their models, only $\sim 13~\%$ of the total amount of seeds formed in protogalaxies
can be SMBH progenitors, because most seeds follow minor branches of the merger tree and 
is in satellite galaxies.
Therefore, the number density we estimated in Eq. (\ref{eq:n_pac}) could be reduced 
by one order of magnitude.
It is worth discussing the occupation fraction of seed BHs in DM halos under streaming velocities
\citep[see also][]{TL14}.

Super-Eddington accretion is an alternative way to form high-z SMBHs with $\ga 10^9~\msun$.
In a massive halo forming a massive seed BH with $M_\bullet \sim 10^5~\msun$, 
the gas density as high as $\ga 10^4~\cc [(1+z)/16]^3$.
In this case, since an HII region formed by ionizing radiation due to the accreting BH 
is confined within the Bondi radius of the BH (i.e., $R_{\rm HII} < R_{\rm Bondi}$),
a steady accretion at $\ga 300~\dot{M}_{\rm Edd}$ can be realized 
(\citealt*{IHO2016,Sakurai_2016}; see also \citealt{Volonteri_Rees_2005,
Alexander_Natarajan_2014,Madau_2014,Pacucci_2015,Sugimura_2017,Sugimura_2018,Takeo_2018})\footnote{
In this paper, the Eddington accretion rate is defined as $\dot{M}_{\rm Edd}\equiv L_{\rm Edd}/c^2$,
and a high accretion rate of $\ga 5000~\dot{M}_{\rm Edd}$ from the Bondi radius is required for 
hyper-Eddington accretion.}, where $\dot{M}_{\rm Edd}\equiv 16~L_{\rm Edd}/c^2$.
A semi-analytical model by \cite{Pezzulli_2016} also supports that hyper-Eddington accretion 
at $\ga 300~\dot{M}_{\rm Edd}$ would likely occur at $10\la z \la 25$.
Moreover, a sufficient amount of gas supply through intense cold streams would feed the BH 
\citep[e.g.,][]{DiMatteo_2012}.
The further study of super(hyper)-Eddington accretion of BHs in more realistic situations
will be left in future work.

We discuss the early stage of massive DM halos hosting seed BHs and PopIII galaxies.
The relation between BHs and galaxies is quite important to understand their coevolution
over the cosmic history \citep{Kormendy_&_Ho_2013}.
From Eqs. (\ref{eq:MBE}) and (\ref{eq:M_disc_pop}), we can estimate the mass ratio of seed BHs to PopIII galaxies as 
$M_\bullet/M_{\star,\rm tot}\sim 0.4~ \epsilon_\star^{-1} ~(T/T_{\rm vir})^{3/2}$, independent of $\theta_{\rm s}$ and $\lambda$.
Since violent mergers of atomic-cooling halos associated with strong BSM can keep the gas pristine 
in halos with $T_{\rm vir}\la 4\times 10^4~\K$, the mass ratio is estimated as $\sim 0.04~\epsilon_\star^{-1}$.
The mass ratio is $\ga 20$ times higher than the SMBH-bulge mass ratio observed 
in the local Universe, $M_\bullet/M_{\rm bulge}\simeq 5\times 10^{-3}(M_{\rm bulge}/10^{11}~\msun)^{0.16}$ 
\citep[e.g.,][]{Magorrian_1998,Gultekin_2009,Kormendy_&_Ho_2013}.
On the other hand, Atacama Large Millimeter/submillimeter Array (ALMA) observations 
of quasar host galaxies at $z\sim 6$
implies that the mass ratios between the SMBHs and the gas dynamical mass are an order of magnitude 
higher than the mean value found in local Universe \citep{Wang_2013,Wang_2016}.
These observational results suggest that SMBHs in high-$z$ quasars remain overmassive.
To explore the SMBH-galaxy coevolution, we need an observational program
with multi-frequencies including near-infrared radiation with JWST and X-rays with a future satellite,
e.g., Lynx\footnote{https://wwwastro.msfc.nasa.gov/lynx/}.

\subsection{Caveats}

\subsubsection{Star formation in cold streams}

Self-gravitating cold streams potentially lead to star formation
\citep{Nakamura_Umemura_2001,Nakamura_Umemura_2002},
preventing shock formation at the edge of the central galactic disc.
In a protogalaxy, the typical line mass density of cold streams is given by
$m_{\rm line}\equiv \pi \rho R_{\rm s}^2=1.1\times 10^4~
T_{\rm vir,4}[(\epsilon_{\rm s}/\epsilon_{\rm v})/0.2]~\msun~\pc^{-1}$.
For a cylindrical filament in a hydro-static equilibrium,
the stability analysis provides a critical line mass density of $m_{\rm crit}=2c_{\rm eff}^2/G$,
above which the filament starts to collapse and fragment due to its self-gravity \citep{Ostriker_1964}.
The effective sound speed is given by $c_{\rm eff}^2=c_{\rm s}^2 + \sigma_{\rm turb}^2$,
where $\sigma_{\rm turb}$ is the turbulent velocity.
One possible origin of the turbulent velocity is BSM\footnote{
Another possible origin of the turbulence would be gas accretion onto cold streams 
from the host halo \citep[e.g.,][]{Klessen_Hennebelle_2010,Heitsch_2013}.
The typical turbulent velocity would be at most $\sim 1~\kms$ in our fiducial case.}.
The typical velocity is roughly estimated as 
$\sigma_{\rm turb} \simeq \sqrt{(v_{\rm cool}^2-v_0^2)}$ or
\begin{equation}
\sigma_{\rm turb} \simeq 3.5~\kms
\left(\frac{v_{\rm bsm}}{2\sigma_{\rm bsm}}\right)
\left(\frac{1+z}{16}\right),
\end{equation}
and thus the critical line mass density is
\begin{align}
m_{\rm crit} & \simeq  5.6\times 10^3~\msun ~\pc^{-1} \nonumber\\
& \times \left[0.56 \left(\frac{T}{10^3~\K}\right) + 
\left(\frac{v_{\rm bsm}}{2\sigma_{\rm bsm}}\right) ^2
\left(\frac{1+z}{16}\right)^2\right].
\end{align}
For $v_{\rm bsm}=2\sigma_{\rm bsm}$, we obtain
$m_{\rm crit }\simeq 8.7\times 10^3 (1.3\times 10^4)~\msun~{\rm pc}^{-1}$ at $z=15 (20)$.
Since the line mass density is marginally higher than the critical value,
the cold stream could be unstable against its self-gravity.
However, we note that this argument is valid for a filamentary stream in hydrostatic equilibrium.
In reality, since a cold stream we consider has a bulk inward supersonic velocity, 
fragmentation would be less efficient than that from a static equilibrium configuration.
Even if the gas filament began to contract and fragment into smaller clumps,
the collapse would proceed in a timescale of $t_{\rm col}=C'/\sqrt{4\pi G\rho}$, where $C'\simeq 3$ 
for the most unstable mode against fragmentation when $m_{\rm line}/m_{\rm crit}\sim 1-2$.
Thus, we obtain the ratio of $t_{\rm col}$ to the infall timescale
of the cold stream $t_{\rm inf}(\equiv r/v_{\rm in})$ as
\begin{align}
\frac{t_{\rm col}}{t_{\rm inf}}=
\frac{C'}{2}\sqrt{\frac{\epsilon_{\rm v}^3}{\epsilon_{\rm s}f_{\rm b}}}
\sim 13
\left(\frac{C'}{3}\right)
\left(\frac{\epsilon_{\rm s}}{0.3}\right)^{-1/2}
\left(\frac{\epsilon_{\rm v}}{1.5}\right)^{3/2}.
\label{eq:inf_ff}
\end{align}
Therefore, we find that the cold stream is unstable against its self-gravity, 
but is likely to accrete onto the central region before fragmenting and forming stars.
Note that the ratio of $m_{\rm line}/m_{\rm crit}$ and Eq. (\ref{eq:inf_ff}) 
do not depend on the choice of $\lambda$.

Recently, \cite{Mandelker_2017} have discussed formation of metal poor globular-clusters
via massive clumps in cold accretion streams in protogalaxies with $M_{\rm halo}\sim 10^{10}~\msun$ at $z\sim 6$.
We note that our scenario does not work in their situation because 
(1) the line mass density tends to be higher in a massive halo and
(2) the critical line mass density becomes smaller at lower redshifts ($z<10$) 
due to the decay of the BSM.
Applying the above argument to a massive halo with $M_{\rm halo}\sim 10^{10}~\msun$ at $z\sim 6$,
we obtain $m_{\rm line}/m_{\rm crit} \ga 10$. 
Since such a massive stream has a higher concentration of the initial density and $C'\la 1$,
the collapse timescale could be comparable or shorter than the infall timescales.

\subsubsection{Shock structure in cold streams}
\label{sec:shock}

In this paper, we discuss the properties of cold accretion flows penetrating deeply 
inside a massive DM halo down to the nuclear disc.
This would be plausible as long as the radiative cooling timescale is shorter than 
the dynamical timescale in the flow, $t_{\rm cool}<t_{\rm dyn}$.
The critical conditions for the deep penetration would be given by
$M_{\rm vir}\la 10^{12}~\msun$ for lower-redshift massive halos
\citep[e.g.,][]{Birnboim_Dekel_2003,Dekel_Birnboim_2006}.
However, the critical conditions in the low mass end are still uncertain.
Here, we assume that deep penetration of cold streams occurs in DM halos
with $T_{\rm vir}\geq 2\times 10^4~\K$ in order to ensure $t_{\rm cool}<t_{\rm dyn}$.
In fact, \cite{Fernandez_2014} found a certain level of penetration even in halos with 
$T_{\rm vir}\simeq 10^4~\K$ when H$_2$ cooling is turned off and Ly$\alpha$ cooling 
carries energy away quickly.
This result supports that what matters is not the gas temperature but the cooling timescale.
They also studied the gas in DM haloes with $T_{\rm vir}\la 10^4~\K$ considering H$_2$ cooling, 
which is less efficient than Ly$\alpha$ cooling.
In this case, cold gas accretes into the halo, but undergoes a series of weak shocks, 
reducing its infall velocity closer to the sound speed.
As a result, the cold stream stalls before reaching the central region, and the shocked density is 
not high enough to dissociate H$_2$.
To explore the nature of cold streams in an atomic-cooling halo and to study the critical conditions 
for deep penetration to form dense shocks are left for future investigations.

For this purpose, we need to investigate a long-term evolution ($\ga$ a few Myr) of a galactic disc 
and cold streams, using some numerical techniques (sink particles and/or a minimum pressure floor) 
to save computation time.
Simultaneously, we have to resolve a small scale of $L$ to capture the physics of interaction 
between the disc and cold streams, and fragmentation of massive clumps at the edge of the disc.
A plausible choice of the minimum spacial length for the simulations 
would be $L< \lambda_{\rm J}/4 \sim 3~\pc$, where $\lambda_{\rm J}$ is the Jeans length at $r\simeq R_{\rm d}$
(see Appendix \ref{sec:length}, also \citealt{Truelove_1997}).

\subsubsection{Dynamical heating or gas collapse triggered by major mergers}
\label{sec:dyn}

In this paper, we assume that major mergers of DM halos disturb a dense gas core 
at the center and thus delay the onset of gravitational collapse of the cloud.
However, one might expect that major mergers trigger the collapse instead.
This issue has been studied by \cite{Chon_2016} in a context of the SMS formation 
aided by strong LW radiation for suppressing H$_2$.
They found that frequent mergers of DM halos generally decrease the gas density at the core 
and prevent gravitational collapse via dynamical heating, which works more efficiently than that in
ordinary PopIII star formation in low-mass DM halos with $\ga 10^5~\msun$ 
\citep[e.g.,][]{Yoshida_2003}.
As discussed in \S\ref{sec:sm}, in this case the size of the gas core is smaller than the local Jeans length, 
and is unlikely to be unstable even after major mergers.
In contrast, \cite{Chon_2016} also found a case where gas collapse can be triggered by an 
{\it extremely frequent} major-merger event
(six gaseous clumps merge in a short timescale; see their Figure 11).
Through the event, the core mass is boosted by a factor of $\sim 30$ only within one dynamical timescale.
Note that mass increase by a factor of $\ga 10$ is required to induce the cloud collapse 
(see their Figure 19b).
This situation seems much more extreme than those we consider in this paper,
but similarly lead to the formation of massive DM halos containing pristine gas.

\subsubsection{External metal enrichment}

We note that external metal enrichment due to supernovae winds
launched from DM halos in metal-polluted trees is negligible.
With strong BSM ($v_{\rm bsm}\ga 2\sigma_{\rm bsm}$), external enrichment 
could be caused by DM halos with $M_{\rm vir}\ga M_{\rm cool}(z)\simeq 2\times 10^6~\msun$ 
at $15<z <25$.
For such halo masses, we require energetic explosions by pair-instability supernovae (PISNe)
to eject heavy elements into intergalactic media (IGM) and cause external metal enrichment
(\citealt{Kitayama_2005,Chiaki_2018}, see also \citealt{Smith_2015,Jeon_2017}). 
Let us suppose that PISNe occur in the majority of DM halos with 
$M_{\rm vir}\ga M_{\rm cool}$ at $z\simeq 20$.
Since the number density of such halos is $n_{\rm h}\simeq  10~\cmpc^{-3}$,
an averaged distance between the PISN halos and pristine DM halos 
due to unusually frequent merger histories can be estimated as $d\simeq 320~\ckpc$ (comoving).
\cite{Greif_2007} have studied a similar situation that a $200~\msun$ PISN occurs 
in an even lower-mass halo with $M_{\rm vir}\simeq 5\times 10^5~\msun$ at $z=20$ and 
enriches the IGM \citep[see also][]{Wise_Abel_2008}.
As a result, a metal-polluted bubble expands only to $\sim 32~\ckpc$ by $z\simeq 15$,
which is much shorter than the typical distance between halos.
Therefore, we can conclude that external metal enrichment of pristine DM halos is unlikely to occur
in a region with $v_{\rm bsm}\ga 2\sigma_{\rm bsm}$.

\section{Summary}
\label{sec:sum}
We propose the formation of massive pristine dark-matter (DM) halos with masses of $\sim 10^8~\msun$, 
due to the dynamical effects by violent mergers in rare regions of the Universe with high baryonic 
streaming velocity relative to DM.
Since the streaming motion prevents gas collapse into DM halos and delays prior star formation episodes,
the gas remains metal-free until the halo reaches virial temperatures $\ga 2\times 10^4~\K$.
The minimum cooling mass of DM halos is boosted by a factor of $\sim 10-30$
because violent major mergers of gaseous halos further inhibit gas collapse.
We use Monte Carlo merger trees to simulate the DM assembly history 
under a streaming velocity of twice the root-mean-square value,
and estimate the number density of massive pristine DM halos as 
$\simeq 10^{-4}~\cmpc^{-3}$.
When the gas infall begins, efficient Ly$\alpha$ cooling drives cold streams penetrating 
inside the halo and feeding a central galactic disc.
When one stream collides with the disc, strong shock forms dense and hot gas cloud, 
where the gas never forms due to effective H$_2$ dissociation.
As a result, a massive gas cloud forms by gravitational instability and collapses
directly into a massive black hole (BH) with $M_\bullet \sim 10^5~\msun$.
Almost simultaneously, a galaxy with $M_{\star, \rm tot}\sim 10^6~\msun$ composed of 
Population III stars forms in the nuclear region.
If the typical stellar mass is as high as $\sim 100~\msun$, the galaxy could be detected 
with the {\it James Webb Space Telescope} even at $z\sim 15$.
The BH would be fed by continuous gas accretion from the host galaxy, and grow to be 
a bright quasar such as those observed at $z\ga 6$.

\section*{Acknowledgements}
We thank Avishai Dekel, Shingo Hirano, Nir Mandelker, Greg Bryan, 
Micheal Norman, Takashi Hosokawa, Kohei Ichikawa, Eli Visbal and Chang-Goo Kim for useful discussions.
This work is partially supported by the Simons Foundation through 
the Simons Society of Fellows (KI) and by NASA grant NNX15AB19G (ZH). 
ML thanks the support the Science Computing Core of Flatiron Institute.

\appendix

\section{Mass accretion rates through cold filaments}
\label{sec:diff_Mdot}

We adopt Eq. (\ref{eq:Mdot_halo}) as gas accretion rates along cold streams.
Here, we discuss a general function form of
\begin{equation}
\dot{M}_{\rm s}\simeq \epsilon_{\rm s}f_{\rm b}A~
T_{\rm vir,4}^{\alpha} \left(\frac{1+z}{16}\right)^{\beta}.
\label{eq:Mdot_halo_app}
\end{equation}
Our fiducial case corresponds to $A=0.41~\msunyr$, $\alpha=3/2$, and $\beta = 0$.
Alternatively, cosmological N-body simulations predict the mean accretion rate of total mass 
onto the virial radius of a massive halo with a mass of $10^{10}\la M_{\rm vir}/\msun \la 10^{15}$ 
at a redshift of $0\leq z<15$ \citep{Fakhouri_2010} 
is well fit by $A=0.16~\msunyr$, $\alpha =1.65$, and $\beta =0.85$.
Note that the later formula might not be applicable for atomic-cooling halos.
The accretion rate in Eq. (\ref{eq:Mdot_halo}) is higher than that in Eq. (\ref{eq:Mdot_halo_app}) 
by a factor of $2.6~(2.0)$ at $z=15~(20)$ for $T_{\rm vir}=10^4~\K$.

\section{Physical properties of galactic discs}
\label{sec:length}

We here summarize the physical length and mass scales on
galactic discs in massive DM halos with $T_{\rm vir}\ga 10^4~\K$.
We adopt a simple model for the vertical structure of the disc, 
which has been used extensively and originally proposed by \cite{Spitzer_1942}.
In this model, where the gas distribution is assumed to be locally isothermal,
the density profile is obtained as
\begin{align}
\rho(r,z)=\rho_0 f(r)~{\rm sech}^2 \left( \frac{z}{2z_{\rm d}}\right),
\end{align}
where $z_{\rm d}=c_{\rm s}/\sqrt{8\pi G \rho_0}$ and $\rho_0$ is the central density,
which is set to the value in Eq. (\ref{eq:rho_disc}).
For this gas profile in the disc, the surface density and the disc mass 
are estimated as $\Sigma_{\rm d}=4\rho_0 z_{\rm d}f(r)$,
and $M_{\rm d}=8\pi I \rho_0 z_{\rm d} R_{\rm d}^2$,
where $I=\int f(r) rdr/R_{\rm d}^2$.
We obtain $I=1$ for an exponential disc, $f(r)=\exp(-r/R_{\rm d})$,
and $I=1/2$ for a uniform disc mode, $f(r)=1$ at $r\leq R_{\rm d}$ and 
$f=0$ at $r>R_{\rm d}$.
In Eq. (\ref{eq:M_disc_pop}), we obtain $M_{\rm d}=2\pi \rho_0 R_{\rm d}^2\times (2z_{\rm d})$,
assuming $I=1/2$ so that the scale height of the disc is given by $2z_{\rm d}$,
\begin{align}
H = \frac{2c_{\rm s}}{V_{\rm vir}}\sqrt{\frac{\epsilon_{\rm v}}{32\epsilon_{\rm s}f_{\rm b}}}\theta_{\rm s}R_{\rm d}
\simeq 0.4~\sqrt{\frac{T}{T_{\rm vir}}}~\hat{\theta}_{\rm s}R_{\rm d},
\label{eq:app1}
\end{align}
where Eq. (\ref{eq:rho_disc}) is used.
Using those equations, the Toomre's parameter is simply given by $Q\simeq \sqrt{H/R_{\rm d}}$.

Finally, let us estimate the Jeans length of the post-shock gas, where 
a massive seed BH could form as discussed in \S\ref{sec:BH}.
Since the gas density behind the shock front is $\ga 4\rho_0$ due to compression by accreting gas and 
atomic-cooling at $T\ga 8000~\K$, the Jeans length is as small as
\begin{align}
\lambda_{\rm J}\la c_{\rm s}\sqrt{\frac{\pi}{4G \rho_0}} \simeq 2 H.
\end{align}
Thus, once dense gas is accumulated at the edge of the galactic disc, 
massive clumps can be unstable and collapse.
Combined with Eq. (\ref{eq:app1}), for $T=8000~\K$ and $T_{\rm vir}=2\times 10^4~\K$, 
the Jeans length is estimated as
\begin{equation}
\lambda_{\rm J}\la 2.5\times 10^{-2}~\hat{\theta_{\rm s}}\hat{\lambda}~R_{\rm vir}
\simeq 13
~\hat{\theta_{\rm s}}\hat{\lambda}~T_{\rm vir,4}^{1/2}
\left(\frac{1+z}{16}\right)^{-3/2}~\pc.
\end{equation}
%

%%%%%%%%%%%%%%%%
\bibliographystyle{mnras}
{
\bibliography{ref}
}

\end{document}